\documentclass[letterpaper,compsoc,twoside]{IEEEtran}
\usepackage{fixltx2e} % LaTeX patches, \textsubscript
\usepackage{cmap} % fix search and cut-and-paste in Acrobat
\usepackage{ifthen}
\usepackage[T1]{fontenc}
\usepackage[utf8]{inputenc}
\usepackage{amsmath}

\usepackage[font={small,it},labelfont=bf]{caption}
\usepackage{float}

\usepackage{textcomp} % text symbol macros

\pdfoutput=1
\usepackage{scipy}
\makeatletter
\def\PY@reset{\let\PY@it=\relax \let\PY@bf=\relax%
    \let\PY@ul=\relax \let\PY@tc=\relax%
    \let\PY@bc=\relax \let\PY@ff=\relax}
\def\PY@tok#1{\csname PY@tok@#1\endcsname}
\def\PY@toks#1+{\ifx\relax#1\empty\else%
    \PY@tok{#1}\expandafter\PY@toks\fi}
\def\PY@do#1{\PY@bc{\PY@tc{\PY@ul{%
    \PY@it{\PY@bf{\PY@ff{#1}}}}}}}
\def\PY#1#2{\PY@reset\PY@toks#1+\relax+\PY@do{#2}}

\expandafter\def\csname PY@tok@gd\endcsname{\def\PY@tc##1{\textcolor[rgb]{0.63,0.00,0.00}{##1}}}
\expandafter\def\csname PY@tok@gu\endcsname{\let\PY@bf=\textbf\def\PY@tc##1{\textcolor[rgb]{0.50,0.00,0.50}{##1}}}
\expandafter\def\csname PY@tok@gt\endcsname{\def\PY@tc##1{\textcolor[rgb]{0.00,0.27,0.87}{##1}}}
\expandafter\def\csname PY@tok@gs\endcsname{\let\PY@bf=\textbf}
\expandafter\def\csname PY@tok@gr\endcsname{\def\PY@tc##1{\textcolor[rgb]{1.00,0.00,0.00}{##1}}}
\expandafter\def\csname PY@tok@cm\endcsname{\let\PY@it=\textit\def\PY@tc##1{\textcolor[rgb]{0.25,0.50,0.56}{##1}}}
\expandafter\def\csname PY@tok@vg\endcsname{\def\PY@tc##1{\textcolor[rgb]{0.73,0.38,0.84}{##1}}}
\expandafter\def\csname PY@tok@vi\endcsname{\def\PY@tc##1{\textcolor[rgb]{0.73,0.38,0.84}{##1}}}
\expandafter\def\csname PY@tok@mh\endcsname{\def\PY@tc##1{\textcolor[rgb]{0.13,0.50,0.31}{##1}}}
\expandafter\def\csname PY@tok@cs\endcsname{\def\PY@tc##1{\textcolor[rgb]{0.25,0.50,0.56}{##1}}\def\PY@bc##1{\setlength{\fboxsep}{0pt}\colorbox[rgb]{1.00,0.94,0.94}{\strut ##1}}}
\expandafter\def\csname PY@tok@ge\endcsname{\let\PY@it=\textit}
\expandafter\def\csname PY@tok@vc\endcsname{\def\PY@tc##1{\textcolor[rgb]{0.73,0.38,0.84}{##1}}}
\expandafter\def\csname PY@tok@il\endcsname{\def\PY@tc##1{\textcolor[rgb]{0.13,0.50,0.31}{##1}}}
\expandafter\def\csname PY@tok@go\endcsname{\def\PY@tc##1{\textcolor[rgb]{0.20,0.20,0.20}{##1}}}
\expandafter\def\csname PY@tok@cp\endcsname{\def\PY@tc##1{\textcolor[rgb]{0.00,0.44,0.13}{##1}}}
\expandafter\def\csname PY@tok@gi\endcsname{\def\PY@tc##1{\textcolor[rgb]{0.00,0.63,0.00}{##1}}}
\expandafter\def\csname PY@tok@gh\endcsname{\let\PY@bf=\textbf\def\PY@tc##1{\textcolor[rgb]{0.00,0.00,0.50}{##1}}}
\expandafter\def\csname PY@tok@ni\endcsname{\let\PY@bf=\textbf\def\PY@tc##1{\textcolor[rgb]{0.84,0.33,0.22}{##1}}}
\expandafter\def\csname PY@tok@nl\endcsname{\let\PY@bf=\textbf\def\PY@tc##1{\textcolor[rgb]{0.00,0.13,0.44}{##1}}}
\expandafter\def\csname PY@tok@nn\endcsname{\let\PY@bf=\textbf\def\PY@tc##1{\textcolor[rgb]{0.05,0.52,0.71}{##1}}}
\expandafter\def\csname PY@tok@no\endcsname{\def\PY@tc##1{\textcolor[rgb]{0.38,0.68,0.84}{##1}}}
\expandafter\def\csname PY@tok@na\endcsname{\def\PY@tc##1{\textcolor[rgb]{0.25,0.44,0.63}{##1}}}
\expandafter\def\csname PY@tok@nb\endcsname{\def\PY@tc##1{\textcolor[rgb]{0.00,0.44,0.13}{##1}}}
\expandafter\def\csname PY@tok@nc\endcsname{\let\PY@bf=\textbf\def\PY@tc##1{\textcolor[rgb]{0.05,0.52,0.71}{##1}}}
\expandafter\def\csname PY@tok@nd\endcsname{\let\PY@bf=\textbf\def\PY@tc##1{\textcolor[rgb]{0.33,0.33,0.33}{##1}}}
\expandafter\def\csname PY@tok@ne\endcsname{\def\PY@tc##1{\textcolor[rgb]{0.00,0.44,0.13}{##1}}}
\expandafter\def\csname PY@tok@nf\endcsname{\def\PY@tc##1{\textcolor[rgb]{0.02,0.16,0.49}{##1}}}
\expandafter\def\csname PY@tok@si\endcsname{\let\PY@it=\textit\def\PY@tc##1{\textcolor[rgb]{0.44,0.63,0.82}{##1}}}
\expandafter\def\csname PY@tok@s2\endcsname{\def\PY@tc##1{\textcolor[rgb]{0.25,0.44,0.63}{##1}}}
\expandafter\def\csname PY@tok@nt\endcsname{\let\PY@bf=\textbf\def\PY@tc##1{\textcolor[rgb]{0.02,0.16,0.45}{##1}}}
\expandafter\def\csname PY@tok@nv\endcsname{\def\PY@tc##1{\textcolor[rgb]{0.73,0.38,0.84}{##1}}}
\expandafter\def\csname PY@tok@s1\endcsname{\def\PY@tc##1{\textcolor[rgb]{0.25,0.44,0.63}{##1}}}
\expandafter\def\csname PY@tok@ch\endcsname{\let\PY@it=\textit\def\PY@tc##1{\textcolor[rgb]{0.25,0.50,0.56}{##1}}}
\expandafter\def\csname PY@tok@m\endcsname{\def\PY@tc##1{\textcolor[rgb]{0.13,0.50,0.31}{##1}}}
\expandafter\def\csname PY@tok@gp\endcsname{\let\PY@bf=\textbf\def\PY@tc##1{\textcolor[rgb]{0.78,0.36,0.04}{##1}}}
\expandafter\def\csname PY@tok@sh\endcsname{\def\PY@tc##1{\textcolor[rgb]{0.25,0.44,0.63}{##1}}}
\expandafter\def\csname PY@tok@ow\endcsname{\let\PY@bf=\textbf\def\PY@tc##1{\textcolor[rgb]{0.00,0.44,0.13}{##1}}}
\expandafter\def\csname PY@tok@sx\endcsname{\def\PY@tc##1{\textcolor[rgb]{0.78,0.36,0.04}{##1}}}
\expandafter\def\csname PY@tok@bp\endcsname{\def\PY@tc##1{\textcolor[rgb]{0.00,0.44,0.13}{##1}}}
\expandafter\def\csname PY@tok@c1\endcsname{\let\PY@it=\textit\def\PY@tc##1{\textcolor[rgb]{0.25,0.50,0.56}{##1}}}
\expandafter\def\csname PY@tok@o\endcsname{\def\PY@tc##1{\textcolor[rgb]{0.40,0.40,0.40}{##1}}}
\expandafter\def\csname PY@tok@kc\endcsname{\let\PY@bf=\textbf\def\PY@tc##1{\textcolor[rgb]{0.00,0.44,0.13}{##1}}}
\expandafter\def\csname PY@tok@c\endcsname{\let\PY@it=\textit\def\PY@tc##1{\textcolor[rgb]{0.25,0.50,0.56}{##1}}}
\expandafter\def\csname PY@tok@mf\endcsname{\def\PY@tc##1{\textcolor[rgb]{0.13,0.50,0.31}{##1}}}
\expandafter\def\csname PY@tok@err\endcsname{\def\PY@bc##1{\setlength{\fboxsep}{0pt}\fcolorbox[rgb]{1.00,0.00,0.00}{1,1,1}{\strut ##1}}}
\expandafter\def\csname PY@tok@mb\endcsname{\def\PY@tc##1{\textcolor[rgb]{0.13,0.50,0.31}{##1}}}
\expandafter\def\csname PY@tok@ss\endcsname{\def\PY@tc##1{\textcolor[rgb]{0.32,0.47,0.09}{##1}}}
\expandafter\def\csname PY@tok@sr\endcsname{\def\PY@tc##1{\textcolor[rgb]{0.14,0.33,0.53}{##1}}}
\expandafter\def\csname PY@tok@mo\endcsname{\def\PY@tc##1{\textcolor[rgb]{0.13,0.50,0.31}{##1}}}
\expandafter\def\csname PY@tok@kd\endcsname{\let\PY@bf=\textbf\def\PY@tc##1{\textcolor[rgb]{0.00,0.44,0.13}{##1}}}
\expandafter\def\csname PY@tok@mi\endcsname{\def\PY@tc##1{\textcolor[rgb]{0.13,0.50,0.31}{##1}}}
\expandafter\def\csname PY@tok@kn\endcsname{\let\PY@bf=\textbf\def\PY@tc##1{\textcolor[rgb]{0.00,0.44,0.13}{##1}}}
\expandafter\def\csname PY@tok@cpf\endcsname{\let\PY@it=\textit\def\PY@tc##1{\textcolor[rgb]{0.25,0.50,0.56}{##1}}}
\expandafter\def\csname PY@tok@kr\endcsname{\let\PY@bf=\textbf\def\PY@tc##1{\textcolor[rgb]{0.00,0.44,0.13}{##1}}}
\expandafter\def\csname PY@tok@s\endcsname{\def\PY@tc##1{\textcolor[rgb]{0.25,0.44,0.63}{##1}}}
\expandafter\def\csname PY@tok@kp\endcsname{\def\PY@tc##1{\textcolor[rgb]{0.00,0.44,0.13}{##1}}}
\expandafter\def\csname PY@tok@w\endcsname{\def\PY@tc##1{\textcolor[rgb]{0.73,0.73,0.73}{##1}}}
\expandafter\def\csname PY@tok@kt\endcsname{\def\PY@tc##1{\textcolor[rgb]{0.56,0.13,0.00}{##1}}}
\expandafter\def\csname PY@tok@sc\endcsname{\def\PY@tc##1{\textcolor[rgb]{0.25,0.44,0.63}{##1}}}
\expandafter\def\csname PY@tok@sb\endcsname{\def\PY@tc##1{\textcolor[rgb]{0.25,0.44,0.63}{##1}}}
\expandafter\def\csname PY@tok@k\endcsname{\let\PY@bf=\textbf\def\PY@tc##1{\textcolor[rgb]{0.00,0.44,0.13}{##1}}}
\expandafter\def\csname PY@tok@se\endcsname{\let\PY@bf=\textbf\def\PY@tc##1{\textcolor[rgb]{0.25,0.44,0.63}{##1}}}
\expandafter\def\csname PY@tok@sd\endcsname{\let\PY@it=\textit\def\PY@tc##1{\textcolor[rgb]{0.25,0.44,0.63}{##1}}}

\def\PYZus{\char`\_}
\def\PYZob{\char`\{}
\def\PYZcb{\char`\}}

\def\PYZlt{\char`\<}
\def\PYZgt{\char`\>}
\def\PYZsh{\char`\#}

\def\PYZsq{\char`\'}
\def\PYZdq{\char`\"}

\makeatother

\providecommand*{\DUrole}[2]{%
  \ifcsname DUrole#1\endcsname%
    \csname DUrole#1\endcsname{#2}%
  \else% backwards compatibility: try \docutilsrole#1{#2}
    \ifcsname docutilsrole#1\endcsname%
      \csname docutilsrole#1\endcsname{#2}%
    \else%
      #2%
    \fi%
  \fi%
}

\providecommand*{\DUroletitlereference}[1]{\textsl{#1}}

\ifthenelse{\isundefined{\hypersetup}}{
  \usepackage[colorlinks=true,linkcolor=blue,urlcolor=blue]{hyperref}
  \urlstyle{same} % normal text font (alternatives: tt, rm, sf)
}{}

\begin{document}
\newcounter{footnotecounter}\title{Using the pyMIC Offload Module in PyFR}\author{Michael Klemm$^{\setcounter{footnotecounter}{1}\fnsymbol{footnotecounter}\setcounter{footnotecounter}{2}\fnsymbol{footnotecounter}}$%
          \setcounter{footnotecounter}{1}\thanks{\fnsymbol{footnotecounter} %
          Corresponding author: \protect\href{mailto:michael.klemm@intel.com}{michael.klemm@intel.com}}\setcounter{footnotecounter}{2}\thanks{\fnsymbol{footnotecounter} Intel Deutschland GmbH, Germany}, Freddie Witherden$^{\setcounter{footnotecounter}{3}\fnsymbol{footnotecounter}}$\setcounter{footnotecounter}{3}\thanks{\fnsymbol{footnotecounter} Imperial College London, UK}, Peter Vincent$^{\setcounter{footnotecounter}{3}\fnsymbol{footnotecounter}}$\thanks{%

          \noindent%
          Copyright\,\copyright\,2015 Intel, Freddie Witherden et al. This is an open-access article distributed under the terms of the Creative Commons Attribution License, which permits unrestricted use, distribution, and reproduction in any medium, provided the original author and source are credited. http://creativecommons.org/licenses/by/3.0/%
        }}\maketitle
          \renewcommand{\leftmark}{PROC. OF THE 8th EUR. CONF. ON PYTHON IN SCIENCE (EUROSCIPY 2015)}
          \renewcommand{\rightmark}{USING THE PYMIC OFFLOAD MODULE IN PYFR}

\setcounter{page}{17}
\newcommand*{\docutilsroleref}{\ref}
\newcommand*{\docutilsrolelabel}{\label}
\AtEndDocument{\cleardoublepage}
\begin{abstract}PyFR is an open-source high-order accurate computational fluid dynamics solver for unstructured grids.
It is designed to efficiently solve the compressible Navier-Stokes equations on a range of hardware platforms, including GPUs and CPUs.
In this paper we will describe how the Python Offload Infrastructure for the Intel Many Integrated Core Architecture (pyMIC) was used to enable PyFR to run with near native performance on the Intel Xeon Phi coprocessor.
We will introduce the architecture of both pyMIC and PyFR and present a variety of examples showcasing the capabilities of pyMIC.
Further, we will also compare the contrast pyMIC to other approaches including native execution and OpenCL.
The process of adding support for pyMIC into PyFR will be described in detail.
Benchmark results show that for a standard cylinder flow problem PyFR with pyMIC is able achieve 240 GFLOP/s of sustained double precision floating point performance; for a 1.85 times improvement over PyFR with C/OpenMP on a 12 core Intel Xeon E5-2697 v2 CPU.\end{abstract}\begin{IEEEkeywords} 
Xeon Phi,
Coprocessor,
Offloading,
CFD\end{IEEEkeywords}

\section{Introduction%
  \label{introduction}%
}

It is a known fact that Python is a programming language that has gained a lot of popularity throughout the computing industry \cite{Tiob14}.
Python is an easy-to-use, elegant scripting language that not only allows for rapid prototyping of ideas, but is also used for the productive development of highly flexible software packages.
Together with NumPy \cite{NumP15}, SciPy \cite{SciP15}, and other packages, Python has been adopted as language for a range of computing problems within the high performance computing (HPC) community.
With these add-on packages, Python can draw from a variety of efficient algorithms that bring Python closer to the performance of compiled languages such as C/C++ and Fortran.

Heterogeneous architectures emerged as a consequence of the desire to compute at a faster pace to shorten time-to-solution or to tackle bigger problem sizes.
Accelerators such as GPGPUs or coprocessors like the Intel® Xeon Phi™ coprocessor \cite{Inte14} are instances of hardware that aim to speed up the floating-point intensive parts of HPC applications.
A typical design involves a cluster of host systems with traditional processors (e.g., Intel® Xeon® processors) that house discrete extension cards.
One usage scenario is the so-called \DUroletitlereference{offload model}, in which the host execution transfers data and control over to the coprocessing device to execute specialized, highly parallel kernels.

In this paper, we present how the Python Offload Infrastructure for the Intel Many Integrated Core Architecture (pyMIC) (see \cite{KlEn14,pyMIC}), a Python module designed to support offloading to the Intel Xeon Phi coprocessor, is used in PyFR \cite{Wit14}.
PyFR is a software package for solving advection-diffusion problems on streaming architectures.
It is designed to solve a variety of governing systems on mixed structured grids consisting of different element types.
Through its execution backends it supports a range of hardware platforms.
A built-in, C-like domain-specific language is used to implement the solver core.
Using the Mako templating engine, the domain-specific language is translated for the backend and execution on the compute system.

The remainder of the paper is organized as follows.
In Section 2, we introduce PyFR and explain its background.
The pyMIC offload module is introduced in Section 3 and Section 4 shows how it has been applied to PyFR.
Section 5 concludes the paper and envisions future work.

\section{Introduction to PyFR%
  \label{introduction-to-pyfr}%
}

PyFR \cite{Wit14} is an open-source Python framework for solving advection-diffusion problems of the form\begin{equation*}
\frac{\partial u}{\partial t} + \nabla \cdot \mathbf{f}(u, \nabla u) = S( \mathbf{x}, t),
\end{equation*}where $u(\mathbf{x},t)$ is a state vector representing the solution, $\mathbf{f}$ a flux function, and $S$ a source term.
A well known example of an advection-diffusion type problem are the compressible Navier-Stokes equations of fluid dynamics.
The efficient solution of which, especially in their unsteady form, is of great interest to both industry and academia.
PyFR is based around the flux reconstruction (FR) approach of Huynh \cite{Huy07}.
FR is both high-order accurate in space and can operate on unstructured grids.
In FR the computational domain of interest is first discretized into a mesh of conforming elements.
Inside of each element two sets of points are defined: one in the interior of the element, commonly termed the \emph{solution points}, and another on the surface of the element, termed the \emph{flux points}.

In FR the solution polynomial inside of each element, as defined by the values of $u$ at the solution points, is discontinuous across elements.
This gives rise to a so-called Riemann problem on the interfaces between elements.
By solving this problem it is possible to obtain a common normal flux polynomial along each interface of an element.
This polynomial can then be used to \emph{correct} the divergence of the discontinuous flux inside of each element to yield an approximation of $\nabla \cdot \mathbf{f}$ that sits in the same polynomial space as the solution.
Once the semi-discretized form has been obtained it can then be used to march the solution forwards in time.
Accomplishing this requires two distinct kinds of operations (i) interpolating/correcting quantities between flux/solution points, and (ii) evaluating quantities (such as the flux) at either individual solution points or pairs of flux points.
When moving quantities, say from the solution points to the flux points, the value at each flux point is given as a weighted sum of the quantity at each solution point\begin{equation*}
q^{(f)}_{e,i} = \sum_j \alpha_{e,ij} q^{(u)}_{e,j},
\end{equation*}where $q$ represents the quantity, $e$ the element number, $i$ the flux point number, and $\alpha_{e,ij}$ is a matrix of coefficients that encodes the numerics.
This can be identified as a matrix-vector product; or in the case of an $N$-element simulation, $N$ matrix-vector products.
If the quantities are first mapped from physical space to a reference space then the $\alpha_{e,ij}$ coefficients become identical for each element of a given type.
Hence, the above operation can be reduced to a single matrix-matrix product.
Depending on the order of accuracy between ${\sim}50\%$ and ${\sim}85\%$ of the wall clock time in an FR code is spent performing such multiplications.
The remaining time is spent in the point-wise operations.
These kernels are a generalization on the form \texttt{f(in1{[}i{]}, in2{[}i{]}, ..., \&out1{[}i{]})}.
As there are no data dependencies between iterations the point-wise kernels are both trivial to parallelize and highly bound by available memory bandwidth.
Given that both matrix multiplication and point-wise evaluation are amenable to acceleration we note
that it is possible to offload all computation within an FR step.
This property is highly desirable as it avoids the need to continually transfer data across the relatively slow PCIe bus.

Our Python implementation of FR, PyFR, has been designed to be compact, efficient, scalable, and performance portable across a range of hardware platforms.
This is accomplished through the use of pluggable \emph{backends}.
Each backend in PyFR exposes a common interface for\newcounter{listcnt0}
\begin{list}{\arabic{listcnt0}.}
{
\usecounter{listcnt0}
\setlength{\rightmargin}{\leftmargin}
}

\item 

memory management;
\item 

matrix multiplication;
\item 

run time kernel generation, compilation, and invocation.\end{list}

Kernels are written \emph{once} in a restrictive C-like domain specific language (DSL) which the backend then translates into the native language of the backend.
In PyFR the DSL is built on top of the popular Mako templating engine \cite{Bay15}.
The specification of the DSL exploits the fact that—at least for point-wise operations—the major parallel programming languages C/OpenMP, CUDA, and OpenCL differ only in how kernels are prototyped and how elements are iterated over.
In addition to portability across platforms the use of a run-time based templating language confers several other advantages.
Firstly, Mako permits Python expressions to be used inside templates to aid in generating the source code for a kernel.
This is significantly more flexible than the C pre-processor and much simpler than C++ templates.
Secondly, as the end result is a Python string it is possible to post-process the code before it is compiled.
A use case for this capability within PyFR is to ensure that when running at single precision that all floating point constants are suffixed by \texttt{.f}.
Doing so helps to avoided unwanted auto-promotion of expressions and avoids the need for awkward casts inside the kernel itself.
Moreover, it is also trivial to allow for user-defined functions and expressions to be inserted into a kernel.
PyFR, for example, permits the form of source term, $S(\mathbf{x},t)$, to be specified as part of the input configuration file.
Without runtime code generation this would require an expression evaluation library and is unlikely to be competitive with the code generated by an optimizing compiler.

An example of a simple kernel written in the DSL can be seen below.%
\begin{quote}\begin{verbatim}
<%inherit file='base'/>
<%namespace module='pyfr.backends.base.makoutil'
  name='pyfr'/>

<%pyfr:kernel name='negdivconf' ndim='2'
        t='scalar fpdtype_t'
        tdivf='inout fpdtype_t[${str(nvars)}]'
        ploc='in fpdtype_t[${str(ndims)}]'
        rcpdjac='in fpdtype_t'>
% for i, ex in enumerate(srcex):
    tdivf[${i}] = -rcpdjac*tdivf[${i}] + ${ex};
% endfor
</%pyfr:kernel>
\end{verbatim}

\end{quote}
There are several points of note.
Firstly, the kernel is purely scalar in nature; choices such as how to vectorize a given operation or how to gather data from memory are all delegated to the backend-specific templating engine.
All the kernel specifies is how to perform a required operation at a single point inside of a single element.
This shields the user from having to understand how data is arranged in memory and permits PyFR to use different memory layouts for different platforms.
Secondly, we note it is possible to utilise Python when generating the main body of kernels.
This capability is used to loop over each of the field variables to generate the body of the kernel.
The variables \texttt{ndims} and \texttt{nvars} refer to the number of spatial dimensions and conservative variables in the system being solved.
It is hence possible to reuse kernels across not only hardware platforms but also governing systems.
Looking at the kernel we observe that two input arguments, \texttt{t} and \texttt{ploc}, appear to go unused.
These correspond to the simulation time $t$ and the physical location $\mathbf{x}$ where the operation is being performed, respectively.
They are potentially referenced by the expressions in \texttt{srcex} which contains a list of source terms to substitute into the kernel body.
During the code generation phase unused arguments are automatically pruned from function prototypes.
This allows PyFR to forego having to allocate memory for $\mathbf{x}$ should the source terms have no spatial dependency.

Currently, backends exist within PyFR for targeting generic CPUs through a C/OpenMP backend, NVIDIA GPUs via a CUDA backend based on PyCUDA \cite{Klö12}, and any device with an OpenCL runtime via an OpenCL backend based on PyOpenCL \cite{Klö12}.
Using these backends PyFR has been shown to be performance portable across a range of platforms \cite{Wit15}.
Sustained performance in excess of 50\% of peak FLOPs has been achieved on both Intel CPUs and NVIDIA GPUs.

To scale out across multiple nodes PyFR has support for distributed memory parallelism using MPI.
This is accomplished through the mpi4py wrappers \cite{Dal15}.
Significant effort has gone into ensuring that communication is overlapped with computation with all MPI requests being both persistent and non-blocking.
Before running PyFR across multiple nodes it is first necessary to decompose the domain using a graph partitioning library such as METIS \cite{Kar98}.
On the Piz Daint supercomputer at CSCS PyFR has been found to exhibit near perfect weak scalability up to 2000 NVIDIA K20X GPUs \cite{Vin15}.
The wire format used by PyFR for MPI buffers is independent of the backend being used.
It is therefore possible for different MPI ranks to use different backends.
This enables simulations to be run on heterogeneous clusters containing a mix of CPUs and accelerators.
However, as discussed in \cite{Wit15}, this capability comes at the cost of a more complicated domain decomposition process.

PyFR v1.0.0 is released under a three-clause new style BSD license and is available from \url{http://pyfr.org}.
The following list summarizes the key functionality of PyFR:%
\begin{itemize}

\item 

Dimensions: 2D, 3D
\item 

Elements: triangles, quadrilaterals, hexahedra, tetrahedra, prisms, pyramids
\item 

Spatial orders: arbitary
\item 

Time steppers: RK4, RK45{[}2R+{]}, TVDRK3
\item 

Precisions: single, double
\item 

Backends: C/OpenMP, CUDA, OpenCL
\item 

Communication: MPI
\item 

File format: parallel HDF5 using h5py \cite{Col13}
\item 

Systems: Euler, compressible Navier-Stokes
\end{itemize}

\section{The pyMIC Module%
  \label{the-pymic-module}%
}

The Python Offload module for the Intel® Many Core Architecture \cite{KlEn14}, follows Python's philosophy by providing an easy-to-use, but widely applicable interface to control offloading to the Intel Xeon Phi coprocessor.
A programmer can start with a very simplistic, maybe non-optimal, offload solution and then refine it by adding more complexity to the program and exercising more fine-grained control over data transfers and kernel invocation.
The guiding principle is to allow for rapid prototyping of a working offload implementation in an application and and then offer the mechanisms to incrementally improve this initial offload solution.
Because NumPy is a well-known and widely used package for (multi-dimensional) array data in scientific Python codes, pyMIC is crafted to blend well with NumPy's \texttt{ndarray} class and its corresponding array operations.

The current version of pyMIC restricts offloaded code to native code for the Intel Xeon Phi coprocessor written in C/C++ or Fortran.
Since most Python codes employ native extension modules for increased execution speed, this blends well with the HPC codes pyMIC is targeting.
Native code can be compiled for the Intel coprocessor and invoked from the Python code through the pyMIC interface.

To foster cross-languge compatibility and to support Python extension modules written in C/C++ and Fortran, pyMIC integrates well with other offload programming models for the Intel coprocessor, such as the Intel® Language Extensions for Offloading (LEO) and the OpenMP 4.0 \texttt{target} constructs.
Programmers can freely mix and match offloading on the Python level with offloading performed in extension modules.
For instance, one could allocate and transfer an \texttt{ndarray} on the Python level through pyMIC's interfaces and then use the data from within an offloaded C/C++ region in an extension module.

\subsection{Architecture%
  \label{architecture}%
}
\begin{figure}[]\noindent\makebox[\columnwidth][c]{\includegraphics[scale=0.60]{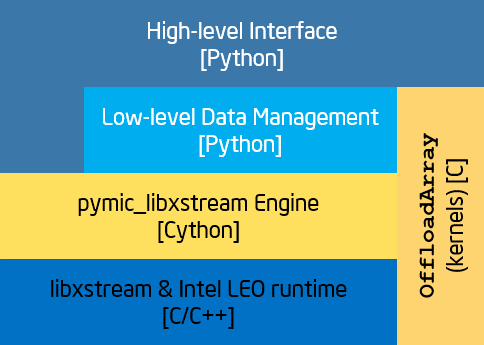}}
\caption{Architecture of the pyMIC offload module. \DUrole{label}{pyMICarch}}
\end{figure}

Figure \DUrole{ref}{pyMICarch} shows the architecture of the pyMIC module.
At the lowest level, the LIBXSTREAM library \cite{Inte15} interacts with the coprocessor devices in the system.
LIBXSTREAM provides a stream-oriented interface to enqueue into an execution stream the invocation of user-defined functions, data allocations, and data transfers.
All enqueued requests are executed asynchronously, but LIBXSTREAM preserves the predecessor/successor relationship of requests within the same stream.
The library is available as open-source software for Intel Architecture.

At the next level up sits the pyMIC offload engine that provides the internal interface for pyMIC's features and that abstracts from the underlying interface of the offload implementation.
This design supports different offload implementations in future versions of pyMIC.
For productivity and easier portability, this level of pyMIC has been implemented in Cython to bridge the gap between the Python level and the LIBXSTREAM library.

The top-level API of pyMIC consists of several classes that provide the different levels of abstractions the offload programming model:%
\begin{itemize}

\item 

\texttt{OffloadDevice} to interact with devices;
\item 

\texttt{OffloadStream} to provide the stream functionality;
\item 

\texttt{OffloadArray} to provide buffer and transfer management;
\item 

and \texttt{OffloadLibrary} for kernel loading and unloading.
\end{itemize}

\subsection{Offloading Code%
  \label{offloading-code}%
}

The following Python code shows how to offload the computation of a \texttt{dgemm} operation to the coprocessor.\begin{Verbatim}[commandchars=\\\{\},numbers=left,firstnumber=1,stepnumber=1,fontsize=\footnotesize,xleftmargin=2.25mm,numbersep=3pt]
\PY{k+kn}{import} \PY{n+nn}{pymic}
\PY{k+kn}{import} \PY{n+nn}{numpy} \PY{k+kn}{as} \PY{n+nn}{np}

\PY{c+c1}{\PYZsh{} size of the matrices}
\PY{n}{m}\PY{p}{,} \PY{n}{n}\PY{p}{,} \PY{n}{k} \PY{o}{=} \PY{l+m+mi}{4096}\PY{p}{,} \PY{l+m+mi}{4096}\PY{p}{,} \PY{l+m+mi}{4096}

\PY{c+c1}{\PYZsh{} create some input data}
\PY{n}{alpha} \PY{o}{=} \PY{l+m+mf}{1.0}
\PY{n}{beta} \PY{o}{=} \PY{l+m+mf}{0.0}
\PY{n}{a} \PY{o}{=} \PY{n}{np}\PY{o}{.}\PY{n}{random}\PY{o}{.}\PY{n}{random}\PY{p}{(}\PY{n}{m}\PY{p}{,} \PY{n}{k}\PY{p}{)}
\PY{n}{b} \PY{o}{=} \PY{n}{np}\PY{o}{.}\PY{n}{random}\PY{o}{.}\PY{n}{random}\PY{p}{(}\PY{n}{k}\PY{p}{,} \PY{n}{n}\PY{p}{)}
\PY{n}{c} \PY{o}{=} \PY{n}{np}\PY{o}{.}\PY{n}{zeros}\PY{p}{(}\PY{n}{m}\PY{p}{,} \PY{n}{n}\PY{p}{)}

\PY{c+c1}{\PYZsh{} load kernel library}
\PY{n}{device} \PY{o}{=} \PY{n}{pymic}\PY{o}{.}\PY{n}{devices}\PY{p}{[}\PY{l+m+mi}{0}\PY{p}{]}
\PY{n}{stream} \PY{o}{=} \PY{n}{device}\PY{o}{.}\PY{n}{get\PYZus{}default\PYZus{}stream}\PY{p}{(}\PY{p}{)}
\PY{n}{library} \PY{o}{=} \PY{n}{device}\PY{o}{.}\PY{n}{load\PYZus{}library}\PY{p}{(}\PY{l+s+s2}{\PYZdq{}}\PY{l+s+s2}{libdgemm.so}\PY{l+s+s2}{\PYZdq{}}\PY{p}{)}

\PY{c+c1}{\PYZsh{} perform the offload and wait for completion}
\PY{n}{stream}\PY{o}{.}\PY{n}{invoke}\PY{p}{(}\PY{n}{library}\PY{o}{.}\PY{n}{mydgemm}\PY{p}{,}
              \PY{n}{a}\PY{p}{,} \PY{n}{b}\PY{p}{,} \PY{n}{c}\PY{p}{,} \PY{n}{m}\PY{p}{,} \PY{n}{n}\PY{p}{,} \PY{n}{k}\PY{p}{,} \PY{n}{alpha}\PY{p}{,} \PY{n}{beta}\PY{p}{)}
\PY{n}{stream}\PY{o}{.}\PY{n}{sync}\PY{p}{(}\PY{p}{)}
\end{Verbatim}
Lines 4-12 initialize the matrix sizes to 4096x4096 elements each and then create two random matrices (\texttt{a}, \texttt{b}) and an empty matrix (\texttt{c}).
Line 15 gets a handle for the first coprocessor of the system and then initializes the default stream to this device (line 16).
Line 17 finally loads a native library that contains the kernel that implements the offloaded version of the \texttt{dgemm} operation.

Lines 19 and 22 enqueue a request to execute the kernel and to synchronize the host thread with the asynchronous kernel invocation.
While the \texttt{invoke} returns immediately after the request has been enqueued into the stream, the \texttt{sync} operation blocks until the kernel execution has finished on the target.

By default, pyMIC provides copy-in/copy-out semantics for the data passed to a kernel.
For NumPy's \texttt{ndarray} objects, the \texttt{invoke} method automatically enqueues allocation and transfer requests from the host to the coprocessor (\DUroletitlereference{copy-in}).
After the request for kernel invocation, corresponding transfers to move data back from the coprocessor are scheduled (\DUroletitlereference{copyout}).
For immutable scalar data, pyMIC only performs the copy-in operation.
While this leads to a very quick first implementation, it also potentially causes unnecessary data transfers.
For instance, although the \texttt{c} matrix is meant to be overwritten on the target (\texttt{beta} is zero), pyMIC would transfer the empty \texttt{c} matrix to the coprocessor and back.
In Section 4.3, we will show how to use pyMIC's interface to optimize data transfers.

The following code example shows the C code of the \texttt{dgemm} kernel:\begin{Verbatim}[commandchars=\\\{\},numbers=left,firstnumber=1,stepnumber=1,fontsize=\footnotesize,xleftmargin=2.25mm,numbersep=3pt]
\PY{c+cp}{\PYZsh{}}\PY{c+cp}{include} \PY{c+cpf}{\PYZlt{}pymic\PYZus{}kernel.h\PYZgt{}}
\PY{c+cp}{\PYZsh{}}\PY{c+cp}{include} \PY{c+cpf}{\PYZlt{}mkl.h\PYZgt{}}

\PY{n}{PYMIC\PYZus{}KERNEL}
\PY{k+kt}{void} \PY{n+nf}{mydgemm}\PY{p}{(}\PY{k}{const} \PY{k+kt}{double} \PY{o}{*}\PY{n}{A}\PY{p}{,} \PY{k}{const} \PY{k+kt}{double} \PY{o}{*}\PY{n}{B}\PY{p}{,}
             \PY{k+kt}{double} \PY{o}{*}\PY{n}{C}\PY{p}{,}
             \PY{k}{const} \PY{k+kt}{int64\PYZus{}t} \PY{o}{*}\PY{n}{m}\PY{p}{,} \PY{k}{const} \PY{k+kt}{int64\PYZus{}t} \PY{o}{*}\PY{n}{n}\PY{p}{,}
             \PY{k}{const} \PY{k+kt}{int64\PYZus{}t} \PY{o}{*}\PY{n}{k}\PY{p}{,}
             \PY{k}{const} \PY{k+kt}{double} \PY{o}{*}\PY{n}{alpha}\PY{p}{,}
             \PY{k}{const} \PY{k+kt}{double} \PY{o}{*}\PY{n}{beta}\PY{p}{)} \PY{p}{\PYZob{}}
     \PY{c+cm}{/* invoke dgemm of MKL\PYZsq{}s cblas wrapper */}
     \PY{n}{cblas\PYZus{}dgemm}\PY{p}{(}\PY{n}{CblasRowMajor}\PY{p}{,} \PY{n}{CblasNoTrans}\PY{p}{,}
                 \PY{n}{CblasNoTrans}\PY{p}{,}
                 \PY{o}{*}\PY{n}{m}\PY{p}{,} \PY{o}{*}\PY{n}{n}\PY{p}{,} \PY{o}{*}\PY{n}{k}\PY{p}{,} \PY{o}{*}\PY{n}{alpha}\PY{p}{,} \PY{n}{A}\PY{p}{,}
                 \PY{o}{*}\PY{n}{k}\PY{p}{,} \PY{n}{B}\PY{p}{,} \PY{o}{*}\PY{n}{n}\PY{p}{,} \PY{o}{*}\PY{n}{beta}\PY{p}{,} \PY{n}{C}\PY{p}{,} \PY{o}{*}\PY{n}{n}\PY{p}{)}\PY{p}{;}
\PY{p}{\PYZcb{}}
\end{Verbatim}
The pyMIC module automatically marshals and unmarshals data that is passed to the offloaded code.
Kernel functions can receive any number of formal parameters, but their signature has to match the actual arguments of the \texttt{invoke} method in the host code.
The types of the formal parameters are pointers to the C/C++ equivalent of a Python scalar type (on Linux*: \texttt{int64\_t}, \texttt{double}, and \texttt{double complex}).
The pointers reference the buffer area that is maintained by pyMIC to keep offloaded data on the coprocessor, so that a kernel can simply access the arguments without calling any additional runtime functions or worrying about data transfers.
However, it is the kernel code's responsibility to access the pointers appropriately and to avoid data corruption when accessing scalar or array data.

In the above \texttt{dgemm} example, the kernel expects the matrices as pointers to data of type \texttt{double}, the matrix sizes as scalar arguments of type \texttt{int64\_t}, and \texttt{alpha} and \texttt{beta} also as pointers to \texttt{double}.
To keep the example simple and to obtain optimal performance, the kernel then invokes the \texttt{dgemm} implementation of the Intel® Math Kernel Library (MKL).

\subsection{Optimizing Data Transfers%
  \label{optimizing-data-transfers}%
}

The following example code shows how to use pyMIC's \texttt{OffloadArray} class to optimize data transfers in the pyMIC programming model.
This can be used to avoid the superfluous data transfers of the above \texttt{dgemm} example.\begin{Verbatim}[commandchars=\\\{\},numbers=left,firstnumber=1,stepnumber=1,fontsize=\footnotesize,xleftmargin=2.25mm,numbersep=3pt]
\PY{k+kn}{import} \PY{n+nn}{pymic}
\PY{k+kn}{import} \PY{n+nn}{numpy} \PY{k+kn}{as} \PY{n+nn}{np}

\PY{c+c1}{\PYZsh{} size of the matrices}
\PY{n}{m}\PY{p}{,} \PY{n}{n}\PY{p}{,} \PY{n}{k} \PY{o}{=} \PY{l+m+mi}{4096}\PY{p}{,} \PY{l+m+mi}{4096}\PY{p}{,} \PY{l+m+mi}{4096}

\PY{c+c1}{\PYZsh{} create some input data}
\PY{n}{alpha} \PY{o}{=} \PY{l+m+mf}{1.0}
\PY{n}{beta} \PY{o}{=} \PY{l+m+mf}{0.0}
\PY{n}{a} \PY{o}{=} \PY{n}{np}\PY{o}{.}\PY{n}{random}\PY{o}{.}\PY{n}{random}\PY{p}{(}\PY{n}{m}\PY{p}{,} \PY{n}{k}\PY{p}{)}
\PY{n}{b} \PY{o}{=} \PY{n}{np}\PY{o}{.}\PY{n}{random}\PY{o}{.}\PY{n}{random}\PY{p}{(}\PY{n}{k}\PY{p}{,} \PY{n}{n}\PY{p}{)}
\PY{n}{c} \PY{o}{=} \PY{n}{np}\PY{o}{.}\PY{n}{zeros}\PY{p}{(}\PY{n}{m}\PY{p}{,} \PY{n}{n}\PY{p}{)}

\PY{c+c1}{\PYZsh{} load kernel library}
\PY{n}{device} \PY{o}{=} \PY{n}{pymic}\PY{o}{.}\PY{n}{devices}\PY{p}{[}\PY{l+m+mi}{0}\PY{p}{]}
\PY{n}{stream} \PY{o}{=} \PY{n}{device}\PY{o}{.}\PY{n}{get\PYZus{}default\PYZus{}stream}\PY{p}{(}\PY{p}{)}
\PY{n}{library} \PY{o}{=} \PY{n}{device}\PY{o}{.}\PY{n}{load\PYZus{}library}\PY{p}{(}\PY{l+s+s2}{\PYZdq{}}\PY{l+s+s2}{libdgemm.so}\PY{l+s+s2}{\PYZdq{}}\PY{p}{)}

\PY{c+c1}{\PYZsh{} create offloaded arrays}
\PY{n}{oa} \PY{o}{=} \PY{n}{stream}\PY{o}{.}\PY{n}{bind}\PY{p}{(}\PY{n}{a}\PY{p}{)}
\PY{n}{ob} \PY{o}{=} \PY{n}{stream}\PY{o}{.}\PY{n}{bind}\PY{p}{(}\PY{n}{b}\PY{p}{)}
\PY{n}{oc} \PY{o}{=} \PY{n}{stream}\PY{o}{.}\PY{n}{bind}\PY{p}{(}\PY{n}{c}\PY{p}{,} \PY{n}{update\PYZus{}device}\PY{o}{=}\PY{n+nb+bp}{False}\PY{p}{)}

\PY{c+c1}{\PYZsh{} perform the offload and wait for completion}
\PY{n}{stream}\PY{o}{.}\PY{n}{invoke}\PY{p}{(}\PY{n}{library}\PY{o}{.}\PY{n}{mydgemm}\PY{p}{,}
              \PY{n}{oa}\PY{p}{,} \PY{n}{ob}\PY{p}{,} \PY{n}{oc}\PY{p}{,} \PY{n}{m}\PY{p}{,} \PY{n}{n}\PY{p}{,} \PY{n}{k}\PY{p}{,} \PY{n}{alpha}\PY{p}{,} \PY{n}{beta}\PY{p}{)}
\PY{n}{oc}\PY{o}{.}\PY{n}{update\PYZus{}host}\PY{p}{(}\PY{p}{)}
\PY{n}{stream}\PY{o}{.}\PY{n}{sync}\PY{p}{(}\PY{p}{)}
\end{Verbatim}
After initializing the data of the matrix as before, the code now uses the \texttt{bind} operation (lines 20 through 22) of the pyMIC API.
The \texttt{bind} operation binds a NumPy \texttt{ndarray} object to an offload buffer of class \texttt{OffloadArray} on the target coprocessor that is associated with a stream object.
The offload buffer is a typed object and contains meta data that descibes the buffer and thus is comparable to a NumPy array.
It also supports basic operations such as element-wise addition, multiplication, zeroing, and filling with values; these operations run as kernels on the coprocessor.
The pyMIC runtime recognizes instances of \texttt{OffloadArray} as kernel arguments and disables automatic copy-in/copy-out transfers for them.

By default the \texttt{bind} operation assumes that the offload buffer should be populated with data from the host array.
To leave the buffer uninitialized and to avoid the data transfer, the \texttt{update\_device} parameter can be set to \texttt{False}.
The \texttt{OffloadArray} instances offer the methods \texttt{update\_device()} and \texttt{update\_host()} enqueue requests for data transfers into the execution stream of the target.
The above example uses this interface to avoid the initial transfer of the \texttt{c} matrix which will be overwritten regardless of its initial values.
In line 27, the code issues an \texttt{update\_host()} call to retrieve the results of the \texttt{mydgemm} kernel.

Where the first example required six data transfers (one copy-in and one copy-out transfers respectively) for \texttt{a}, \texttt{b}, and \texttt{c}, the last example only performs the minimal number of transfers, that is, it transfers \texttt{a} and \texttt{b} from the host to the device and only moves \texttt{c} back to the host process.

\subsection{The pyMIC Low-level Interface%
  \label{the-pymic-low-level-interface}%
}

PyFR's offload model needs more fine-grained control over memory management and referencing data on the target device.
While such low-level interactivity enables the programmer to exercise full control over all aspects of the offload workflow, it also exposes a lot of details such as device pointers and memory offsets.
The low-level data management interface (see Figure \DUrole{ref}{pyMICarch}) that pyMIC uses internally is therefore intentionally exposed as part of the pyMIC API.

This interface is based on \texttt{memcpy}-like methods of a device stream.
It supports allocation and deallocation of \texttt{nbytes} of device data with a given data aligment:\begin{Verbatim}[commandchars=\\\{\},fontsize=\footnotesize]
\PY{n}{allocate\PYZus{}device\PYZus{}memory}\PY{p}{(}\PY{n+nb+bp}{self}\PY{p}{,} \PY{n}{nbytes}\PY{p}{,} \PY{n}{alignment}\PY{o}{=}\PY{l+m+mi}{64}\PY{p}{)}
\PY{n}{deallocate\PYZus{}device\PYZus{}memory}\PY{p}{(}\PY{n+nb+bp}{self}\PY{p}{,} \PY{n}{device\PYZus{}ptr}\PY{p}{)}
\end{Verbatim}
It also offers primitive operations for different directions of data transfers:\begin{Verbatim}[commandchars=\\\{\},fontsize=\footnotesize]
\PY{n}{transfer\PYZus{}host2device}\PY{p}{(}\PY{n+nb+bp}{self}\PY{p}{,} \PY{n}{host\PYZus{}ptr}\PY{p}{,} \PY{n}{device\PYZus{}ptr}\PY{p}{,}
                     \PY{n}{nbytes}\PY{p}{,}
                     \PY{n}{offset\PYZus{}host}\PY{o}{=}\PY{l+m+mi}{0}\PY{p}{,} \PY{n}{offset\PYZus{}device}\PY{o}{=}\PY{l+m+mi}{0}\PY{p}{)}
\PY{n}{transfer\PYZus{}device2host}\PY{p}{(}\PY{n+nb+bp}{self}\PY{p}{,} \PY{n}{device\PYZus{}ptr}\PY{p}{,} \PY{n}{host\PYZus{}ptr}\PY{p}{,}
                     \PY{n}{nbytes}\PY{p}{,}
                     \PY{n}{offset\PYZus{}device}\PY{o}{=}\PY{l+m+mi}{0}\PY{p}{,} \PY{n}{offset\PYZus{}host}\PY{o}{=}\PY{l+m+mi}{0}\PY{p}{)}
\PY{n}{transfer\PYZus{}device2device}\PY{p}{(}\PY{n+nb+bp}{self}\PY{p}{,}
                       \PY{n}{device\PYZus{}ptr\PYZus{}src}\PY{p}{,}
                       \PY{n}{device\PYZus{}ptr\PYZus{}dst}\PY{p}{,}
                       \PY{n}{nbytes}\PY{p}{,}
                       \PY{n}{offset\PYZus{}device\PYZus{}src}\PY{o}{=}\PY{l+m+mi}{0}\PY{p}{,}
                       \PY{n}{offset\PYZus{}device\PYZus{}dst}\PY{o}{=}\PY{l+m+mi}{0}\PY{p}{)}
\end{Verbatim}
Similar to the high-level interface of pyMIC, it's low-level interface operates using a stream-based model.
All of the above methods may be executed asynchronously and require to call the \texttt{sync} operation to wait for completion.

The host pointer passed as an argument is an actual pointer as returned by NumPy's \texttt{nadrray.ctypes.data} or similar operations that expose a C-style pointer into the host memory associated with a Python object.
The device pointer is a fake pointer that was returned by \texttt{allocate\_device\_memory} and that uniquely identifies the data allocation on the target device.
Note that these allocations are smart in the sense that once the Python garbage collector reclaims a smart pointer, the \texttt{\_\_del\_\_} method automatically releases the device memory associated with the allocation.

\section{Using pyMIC to Offload PyFR%
  \label{using-pymic-to-offload-pyfr}%
}

Although PyFR can be run on the Intel Xeon Phi coprocessor using the OpenCL backend this configuration is not optimal.
As was outlined in Section 2 the performance of PyFR depends heavily on the presence of a highly tuned matrix multiplication library.
For the coprocessor this is the Intel MKL.
However, as the MKL does not provide an OpenCL interface it is necessary to implement these kernels using pure OpenCL code.
This is known to be a challenging problem \cite{McI14}.
Hence, in order to take full advantage of the capabilities of the coprocessor a native approach is required.

One possible approach here is to move PyFR in its entirety onto the Phi itself and then run with the C/OpenMP backend.
However, this requires that Python, along with dependencies such as NumPy, be cross-compiled for the Intel coprocessor; a significant undertaking.
Additionally, as the Intel compiler does not run natively on the coprocessor an additional set of scripts would also be required to ‘offload’ the compilation of runtime-generated kernels onto the host.
Moreover, with this approach the initial start up phase would also be run on the coprocessor.
As the single-thread performance of the Intel Xeon Phi coprocessor is significantly less than that of a recent Xeon processor, this is likely to result in a substantial increase in the start-up time of PyFR.
Trying to compensate for this additional overheads might render the native solution ineffective.
It was therefore decided to add a native MIC backend into PyFR and do so by leveraging pyMIC.

On account of its need to target CUDA* and OpenCL the PyFR backend interface is relatively low-level.
At start up, the solver code in PyFR allocates large blocks of memory which it then slices up into smaller pieces.
A backend must therefore provide a means of both allocating memory and copying regions of this memory to/from the host.
In contrast to this pyMIC is a relatively high-level library whose core tenant is comparable to a NumPy's \texttt{ndarray} type.
While writing the MIC backend for PyFR it was therefore necessary to use the low-level interfaces to pyMIC that enables raw memory to be allocated on the device and fine-grained copying to/from this memory.

The resulting backend consists of approximately 700 lines of pure Python code and 200 lines of Mako templates.
As the native programming language for the Intel coprocessor is C code with OpenMP annotations the DSL translation engine for the Intel coprocessor is almost identical to the one used in the existing C/OpenMP backend with the only changes being around how arguments are passed into kernels.
These generated kernels are then compiled at runtime by invoking the Intel compiler on the host to produce a shared library.
The PyFR framework then loads the library on the target device by executing the \texttt{load\_library} method of the device handle.

Matrix multiplications are handled by invoking a native kernel which itself calls out to the \texttt{cblas\_sgemm} and \texttt{cblas\_dgemm} routines from MKL.
This provides the optimal implementation to execute matrix multiplies on the coprocessor.

\section{Performance Results%
  \label{performance-results}%
}

To evaluate the performance of PyFR with pyMIC as an execution backend, a system with an Intel Xeon E5-2697 v2 host process and a Intel Xeon Phi 3120A coprocessor was employed.  However, before evaluating the performance of PyFR with pyMIC it is first useful to consider the raw, standalone, performance of the pyMIC module.

\subsection{Performance of pyMIC%
  \label{performance-of-pymic}%
}
\begin{figure}[]\noindent\makebox[\columnwidth][c]{\includegraphics[scale=0.60]{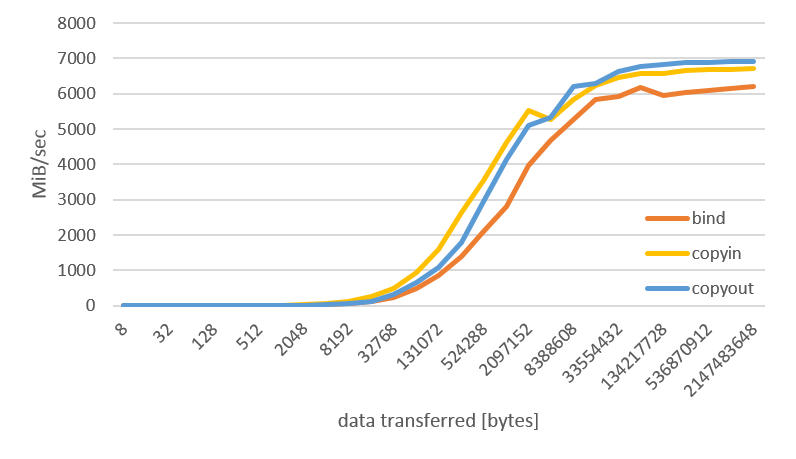}}
\caption{Bandwidth of the data-transfer operations of pyMIC (see \cite{KlEn14}). \DUrole{label}{pyMICPerfBandwidth}}
\end{figure}\begin{figure}[]\noindent\makebox[\columnwidth][c]{\includegraphics[scale=0.60]{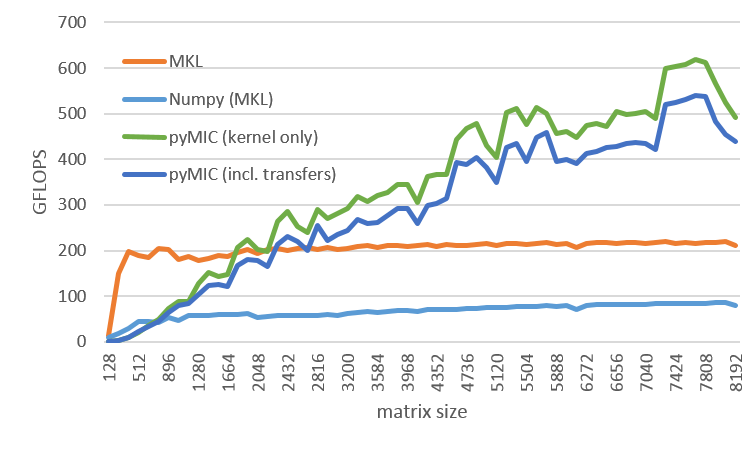}}
\caption{Performance of the offloadded \texttt{dgemm} operation(see \cite{KlEn14}). \DUrole{label}{pyMICPerfDgemm}}
\end{figure}

Figures \DUrole{ref}{pyMICPerfBandwidth} shows the performance results of micro-benchmarks that measure the achieved bandwidth as reported in \cite{KlEn14}.
The achieved bandwidth depends on the size of the data transfer.
For short data transfers, latency of enqueuing the request and setting up the data transfer in the offload runtime dominates, so that the achieved bandwidth is low.
With increasing transfer size, latency becomes less important and thus bandwidth goes up until it saturates at the PCIe gen2 limit.
The effective bandwidth of the bind operation is lower, because it involves the overhead of allocation of the offload buffer, while pure transfers (\DUroletitlereference{copyin} and \DUroletitlereference{copyout}) move data into existing buffers.

Figure \DUrole{ref}{pyMICPerfDgemm} depicts the GFLOPS rate of offloading the \texttt{dgemm} operation (cf. \cite{KlEn14}).
The chart compares the MKL native \texttt{dgemm} operation of a micro-benchmark written in C (\DUroletitlereference{MKL}) with the performance of NumPy that was setup to use MKL (\DUroletitlereference{NumPy (MKL)}).
Both are executing on the host for various quadratic matrix sizes as our baseline.
The chart also shows the \texttt{mydgemm} kernel comparing \DUroletitlereference{pyMIC (kernel only)} and \DUroletitlereference{pyMIC (incl. transfers)}.
As can be seen the GFLOPS rate of MKL quickly saturates at small matrix sizes because of the effective threading implementation used.
Due to the cache-blocking in MKL, it provides a stable level of performance across all matrix sizes once it has saturated.
The comparatively low performance of NumPy is attributed to several temporary copies that NumPy has to maintain to implement a full \texttt{dgemm} operation.
Offloading the kernel for small matrix sizes is not expected to yield any performance gain due to the latency of transferring the small matrices from the host to the coprocessor.
For matrices larger than 2048x2048 elements, the coprocessor is able to compensate the transfer latency and to yield better performance than the host system.
Naturally, the effective GFLOP rate is slightly lower if data transfers are taken into consideration.

\subsection{Performance of PyFR%
  \label{performance-of-pyfr}%
}

As a benchmark problem we consider the case of flow over a circular cylinder at Mach 0.2 and Reynolds number 3900.
Following \cite{Wit14} the domain was meshed with 46610 hexahedra and run with fourth order solution polynomials.
A visual depiction of the simulation can be seen in Figure \DUrole{ref}{pyfrcyl}.
When running at double precision this gives a working set of 3.1 GiB.
One complete time step using a fourth order Runge-Kutta scheme requires on the order of ${\sim}4.6 \times 10^{11}$ floating point operations with typical simulations requiring on the order of half of million steps.
The performance of PyFR in sustained GFLOPS for this problem on an Intel Xeon Phi 3120A coprocessor (57 cores at 1.1 GHz) can be seen in Figure \DUrole{ref}{pyfrperf}.
Results for a twelve core Intel Xeon E5-2697 v2 CPU using the OpenMP backend are also included.
Using pyMIC a speedup of approximately 1.85 times can be observed.
Further, 11 of the CPU cores are freed up in the process to run either alternative workloads or a heterogenous PyFR simulation using two MPI ranks to exploit both the CPU cores and the coprocessor.\begin{figure}[]\noindent\makebox[\columnwidth][c]{\includegraphics[scale=0.12]{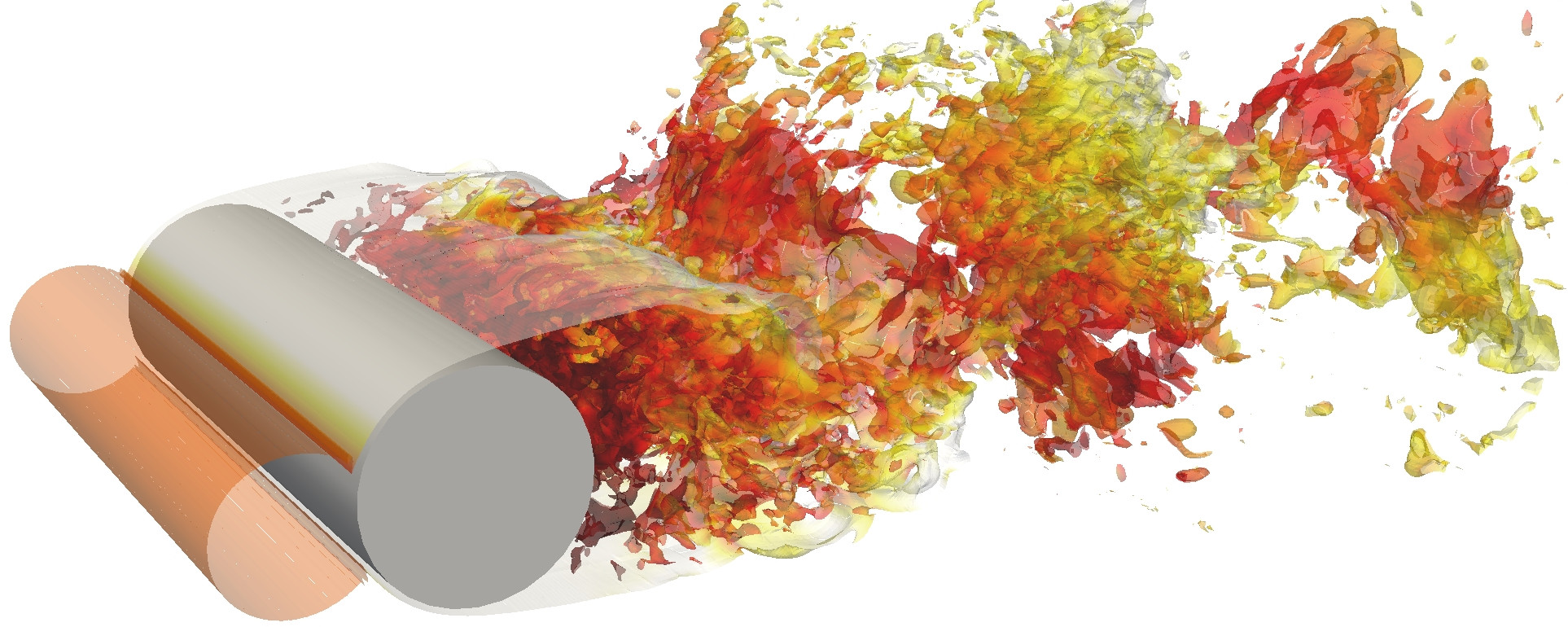}}
\caption{Isosurfaces of density colored by velocity magnitude for the cylinder benchmark problem. \DUrole{label}{pyfrcyl}}
\end{figure}\begin{figure}[]\noindent\makebox[\columnwidth][c]{\includegraphics[scale=0.60]{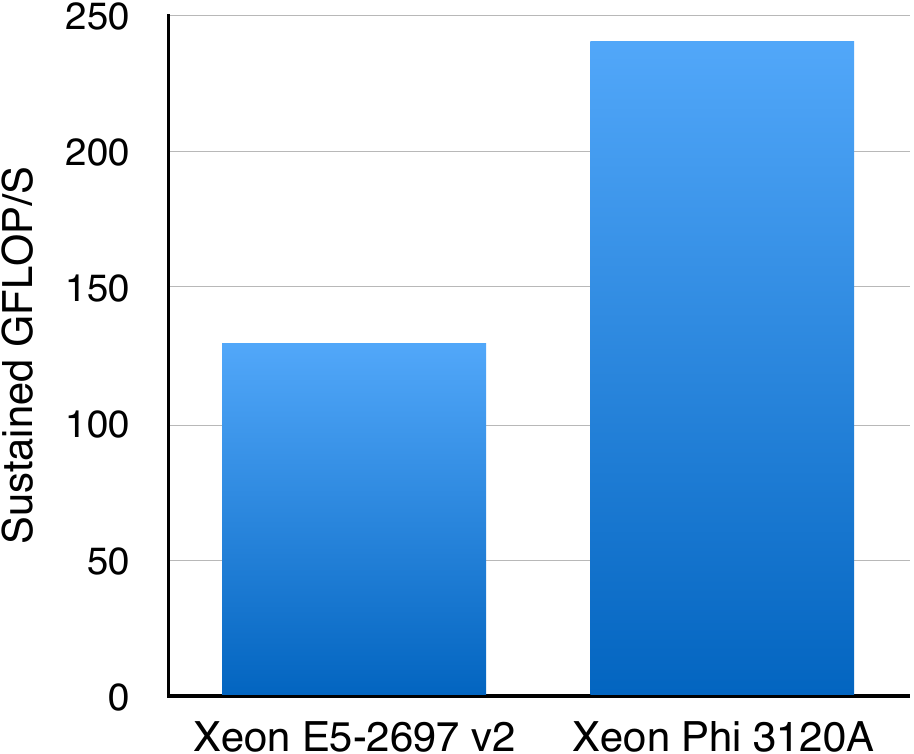}}
\caption{Sustained performance of PyFR for the cylinder flow problem using the C/OpenMP backend on a 12 core Xeon E5-2697 CPU and the pyMIC backend on an actively cooled Xeon Phi 3120A. \DUrole{label}{pyfrperf}}
\end{figure}

\section{Conclusion and Future Work%
  \label{conclusion-and-future-work}%
}

In this paper we have introduced the pyMIC offload module for executing kernels on the Intel Xeon Phi coprocessor.
The architecture of pyMIC has been outlined and several examples have been presented.
It is shown by utilizing pyMIC in combination with MKL how it is possible to obtain a substantial speedup for \texttt{dgemm}.
We have also described PyFR, an open source framework for solving the compressible Navier-Stokes equations.
The architecture of PyFR, including the techniques used that allow it to run performantly across a variety of hardware platforms, have also been presented.
We have shown how using pyMIC it is possible add a backend into PyFR that can target the Intel Xeon Phi coprocessor.
Implementation details have been discussed and benchmarks presented that show a speedup compared with a conventional CPU for a benchmark flow problem.

The roadmap for the pyMIC module contains several extensions that we are planning to develop over the course of the upcoming releases.
The next release of pyMIC will support Python 3.
We are also working on extending the synchronization capabilities of pyMIC to
add support for multiple independent streams.
A future version of pyMIC will add events that will allow for synchronizing host threads with streams objects as well as the synchronization of multiple streams.
Compression of the data stream to the target device may be of interest to leverage the compute power of the host system to compress the data stream before sending it over the PCIe bus.
Finally, we are looking into extending pyMIC beyond native kernels on the target devices by providing offload capabilities for generic Python code.

\section{Acknowledgments%
  \label{acknowledgments}%
}

Peter Vincent and Freddie Witherden would like to thank the Engineering and Physical Sciences Research Council for their support via a Doctoral Training Grant, an Early Career Fellowship (EP/K027379/1), and the Hyper Flux project (EP/M50676X/1).

Intel, Xeon, and Xeon Phi are trademarks or registered trademarks of Intel Corporation or its subsidiaries in the United States and other countries.

* Other names and brands are the property of their respective owners.

Software and workloads used in performance tests may have been optimized for performance only on Intel microprocessors.
Performance tests, such as SYSmark and MobileMark, are measured using specific computer systems, components, software, operations and functions.
Any change to any of those factors may cause the results to vary.
You should consult other information and performance tests to assist you in fully evaluating your contemplated purchases, including the performance of that product when combined with other products.
For more information go to \url{http://www.intel.com/performance}.

Intel's compilers may or may not optimize to the same degree for non-Intel microprocessors for optimizations that are not unique to Intel microprocessors.
These optimizations include SSE2, SSE3, and SSSE3 instruction sets and other optimizations.
Intel does not guarantee the availability, functionality, or effectiveness of any optimization on microprocessors not manufactured by Intel. Microprocessor-dependent optimizations in this product are intended for use with Intel microprocessors.
Certain optimizations not specific to Intel microarchitecture are reserved for Intel microprocessors.
Please refer to the applicable product User and Reference Guides for more information regarding the specific instruction sets covered by this notice.

\end{document}